\documentclass{optica-article}

\journal{opticajournal} 

\articletype{Research Article}

\UseRawInputEncoding
\usepackage{lineno}
\usepackage{float}
\usepackage{graphicx}

\begin{document}

\title{Phase-controlled super-modes in a `lossy' tri-waveguide coupled micro-ring resonator system}

\author{Tianrui Li, Max Goodwin, Matthew P. Halsall and Iain F. Crowe *}

\address{Department of Electrical and Electronic Engineering, Photon Science Institute, The University of Manchester, Manchester, M13 9PL, UK;}

\email{\authormark{*}iain.crowe@manchester.ac.uk} 


\begin{abstract*} 
In coupled micro-ring resonator (MRR) systems, the coupling process plays a critical role in determining the optical response, yet models of these structures typically assume this process to be either lossless, or dissipative with either purely real or purely imaginary coupling rate, respectively. Dissipative coupling is contingent on having appropriately phase engineered structures with intricate coupling geometries, but systems exhibiting `lossy coupling', i.e., characterized by complex-valued coupling rates, have received comparatively little attention. In this work, we investigate the effect of such `lossy' coupling, by modelling the optical response of a tri-waveguide coupled MRR system, using Temporal Coupled-Mode Theory (TCMT). The `lossy' coupling occurs via a common bus waveguide, positioned between a pair of MRRs, leading to the emergence of a pair of intrinsic resonances, whose frequency and linewidth may be tailored by controlling the respective coupling rates from either MRR. We develop these ideas, and by connecting the input (driving) waveguides, using a 3dB y-splitter, demonstrate a novel, coherently driven, lossy-coupled MRR, which exhibits either an absorptive or transparent resonance peak on demand by phase-controlled super-modes.
\end{abstract*}

\section{Introduction}
In the rapidly evolving field of integrated photonics, the demand for compact and efficient optical devices has driven the exploration of new resonator architectures, including photonic crystal (PhC) based  micro-ring resonators (MRRs) \cite{wang_optimizing_2025}, and MRRs integrated with 2D materials \cite{hussein_raman_2017,crowe_determination_2014}. Standard, waveguide based MRRs already play a central role in many photonic integrated circuits (PICs) because of their versatility and broad functionality, including laser mode locking, optical modulation, spectral filtering, and sensing \cite{midya_supersymmetric_2019,longhi_unidirectional_2017,yang_efficient_2022,dingel_ultra-linear_2004,zhang_bandwidth_2022,lu_parity-time_2022}. Combining MRRs with a Mach-Zehnder interferometer, in so-called ring-enhanced MZI (REMZI) structures, can further enhance sensitivity and tunable filtering \cite{dai_highly_2009,ding_bandwidth_2011,zheng_integrated_2021}.

Other such novel PIC architectures include coupled MRRs for implementation of `on-chip' non-Hermitian photonics, and parity-time (PT) symmetry\cite{peng_paritytime-symmetric_2014,li_n-order_2024}, providing an alternative route to improved performance \cite{zhao_paritytime_2018}. A similar approach employs what is referred to as phase-controlled super-modes \cite{boriskina_symmetry_2005, abbaslou_high-spectral-contrast_2016,geints_phase-controlled_2023} to selectively induce or suppress the symmetric (S) or antisymmetric (AS) mode(s) of a coupled MRR, offering yet another degree of control for such device applications.

Analysis of such devices requires proper treatment of the inter-ring coupling, which is often assumed to be lossless, i.e., Hermitian conjugate. In practice, however, losses arising during the inter-ring coupling process are usually treated as perturbations or fabrication imperfections rather than as intrinsic elements of the system \cite{nag_chowdhury_exceptional-point-enhanced_2025}. Furthermore, these coupling losses are difficult to quantify directly in either simulations or experiments, which makes them challenging to incorporate into idealized theoretical models.

A useful platform for investigating this problem is the tri-waveguide coupled-MRR system \cite{chen_all-optical_2013,zhou_wavelength-selective_2015}. In this structure, the loss introduced during the coupling process ideally corresponds to radiation through a common (coupling) waveguide. This basic configuration has already revealed several interesting features. For example, Chen Shen and co-workers employed such a structure to realize photonic analogues of Kirchhoff's law, and numerical solutions of Laplace's equation \cite{sun_induced_2021,ye_reconfigurable_2024}. However, those studies relied primarily on numerical optimization and did not establish a general theoretical framework.

Although the transfer-matrix method (TMM) has been deployed to model such structures\cite{zhou_wavelength-selective_2015,cherchi_3x3_2018}, this is less suitable for analyzing non-Hermitian systems, compared with, say, temporal coupled-mode theory (TCMT) \cite{van_optical_2016,behunin_fundamental_2018,hu_-chip_2021}. However, even TCMT \cite{de_carlo_anti-pt-symmetric_2023} is not always consistent with TMM \cite{zhou_wavelength-selective_2015}, motivating the need for a rigorous, self-consistent TCMT description of such systems.

In this paper, we develop a TCMT-based framework for a tri-waveguide coupled MRR system, specifically with a complex-valued inter-ring coupling rate that fully captures the `lossy' coupling process. Our model reveals an optical response that differs fundamentally from those of both conventional Hermitian, and purely dissipative, coupling. We implement these ideas by developing a system in which the outer waveguides of the tri-waveguide coupler are connected to a common input, using a 3dB y-splitter, enabling a phase-controlled super-modes system that exhibits absorption resonance on demand.

\section{`Lossy' coupling: theoretical formulation}
From a quantum-mechanical perspective, an MRR operating at a fixed resonance frequency may be modelled as a harmonic oscillator. When two such resonators are coupled and driven by an external optical field, the resulting system can be described by the Hamiltonian of two coupled harmonic oscillators: 

\begin{equation}
    \label{equ:6-1}
    H = \omega_1 a_1^\dagger a_1 + \omega_2 a_2^\dagger a_2 + \mu_a \left( a_1^\dagger a_2 + a_1 a_2^\dagger \right) + \mu_b \, a_{in} \, a_1^\dagger
\end{equation}

Here, $\omega_1$, $\omega_2$ are the resonant frequencies of the two resonators, $a_{i}^{\dagger}$ and $a_i$ are the raising and lowering operators and $\mu_{a}$ is the coupling strength between the two resonators. The final term, $\mu_b \, a_{in} \, a_1^\dagger$ represents the coupling of the external (driving) field into the first resonator.

Substituting this Hamiltonian into the Heisenberg equation yields the dynamical equations for the two resonators:

\begin{equation}
    \label{equ:6-2}
    j \frac{d}{dt} a_1 = \left[ a_1,\,\omega_1 a_1^\dagger a_1 + \omega_2 a_2^\dagger a_2
+ \mu_a \left( a_1^\dagger a_2 + a_1 a_2^\dagger \right)
+ \mu_b \, a_{in} \, a_1^\dagger \right]
\end{equation}
\begin{equation}
    \label{equ:6-3}
    j \frac{d}{dt} a_2 = \left[ a_2,\,\omega_1 a_1^\dagger a_1 + \omega_2 a_2^\dagger a_2
+ \mu_a \left( a_1^\dagger a_2 + a_1 a_2^\dagger \right)
+ \mu_b \, a_{in} \, a_1^\dagger \right]
\end{equation}

And expanding the commutators gives the following expressions:

\begin{equation}
    \label{equ:6-4}
    j \frac{d}{dt} a_1 = \omega_1 a_1 + \mu_a a_2 + \mu_b a_{in}
\end{equation}
\begin{equation}
    \label{equ:6-5}
    j \frac{d}{dt} a_2 = \omega_2 a_2 + \mu_a a_1
\end{equation}

According to the quantum Langevin formalism, coupling between each resonator and the external environment can be regarded as a `lossy' process \cite{milburn_quantum_2015}. Consequently, the resonant frequencies in Eqs. \ref{equ:6-4} and \ref{equ:6-5}, $\omega_i$ are replaced by complex-valued frequencies, $\omega_i-j\gamma_i$, with $\gamma_i$ the intrinsic loss rate of the $i^{th}$ resonator.

Similarly, if the inter-ring coupling is treated as a `lossy' process, i.e., if some energy is lost to the environment during coupling via the central waveguide, then $\mu_a \rightarrow \mu_a-j\gamma_a$ with $\gamma_a$ the associated coupling loss.

In the quantum mechanical framework, $a_i$ are operators acting on the wavefunction to yield physical observables. Consequently, from an electrodynamic perspective, these correspond to the intracavity field amplitudes of the two resonators. Combining Eqs. \ref{equ:6-4} and \ref{equ:6-5}, the temporal coupled-mode equations of the `lossy' coupled-MRR system can therefore be written in a single matrix form:

\begin{equation}
    \label{equ:6-6}
    j \frac{d}{dt}
    \begin{bmatrix}
    a_1 \\
    a_2
    \end{bmatrix}
    =
    \begin{bmatrix}
    \omega_1 - j\gamma_1 & \mu_a - j\gamma_a \\
    \mu_a - j\gamma_a & \omega_2 - j\gamma_2
    \end{bmatrix}
    \begin{bmatrix}
    a_1 \\
    a_2
    \end{bmatrix}
    +
    \begin{bmatrix}
    \mu_b \\
    0
    \end{bmatrix}
    a_{in}
\end{equation}

And, if we write:

\begin{equation}
    \notag
    \psi =
    \begin{bmatrix}
    a_1 \\
    a_2
    \end{bmatrix} , \quad 
    H =
    \begin{bmatrix}
    \omega_1 - j\gamma_1 & \mu_a - j\gamma_a \\
    \mu_a - j\gamma_a & \omega_2 - j\gamma_2
    \end{bmatrix}
\end{equation}

Eq. \ref{equ:6-6} takes the form of the time-dependent Schr{\"o}dinger equation, $j\frac{d}{dt}\psi=H\psi+\left[\begin{matrix}\mu_b\\0\end{matrix}\right]a_{in}$ with an external driving term, $\left[\begin{matrix}\mu_b\\0\end{matrix}\right]a_{in}$. In this formulation, the eigenvalues of the Hamiltonian, H determine the resonant modes of the coupled system, with the resonant frequency (linewidth) given by the real (imaginary) parts.

Previous studies of dissipative coupling \cite{yang_anti-mathcalpt_2017,li_experimental_2024} have mainly considered the special case in which the coupling coefficient is purely imaginary, i.e., $\mu_a=0$. In contrast, the present work considers the more general case of genuinely complex-valued coupling. To distinguish this regime from purely dissipative coupling, we refer to it here as `lossy' coupling.

\section{Tri-waveguide implementation of `lossy' coupling}

\begin{figure}[h]
    \centering
    \includegraphics[width=0.5\linewidth]{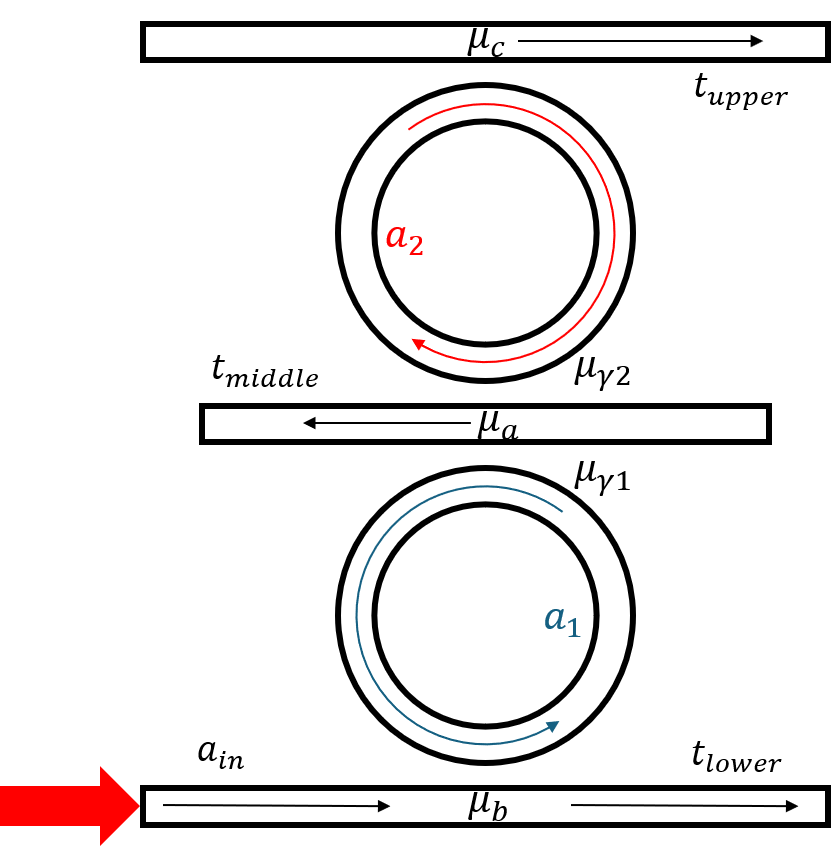}
    \caption{Schematic of the tri-waveguide `lossy' coupled MRR system, driven with an input optical signal at the bottom left (red arrow)}
    \label{fig:6-1}
\end{figure}

Based on the preceding analysis, achieving a complex-valued coupling coefficient requires loss during the coupling process rather than phase by engineering alone. Here, we employ a simple tri-waveguide coupler to realize `lossy' coupling via a common waveguide positioned between two MRRs, as illustrated in Fig. \ref{fig:6-1}.

Although the structure in Fig. \ref{fig:6-1} has previously been analyzed using $2 \times 2$ and $3 \times 3$ TMMs \cite{zhou_wavelength-selective_2015,cherchi_3x3_2018}, those approaches involve lengthy calculations and do not clearly reveal the non-Hermitian character of such systems through eigenvalue analysis. In the idealized case considered here, the coupling between the input/output waveguides and the MRRs is assumed to be lossless. Consequently, any loss associated with inter-ring coupling is assumed to radiate entirely through the common, central waveguide.

In Fig. \ref{fig:6-1}, the coupling rates between waveguides and MRRs are defined according to light entering the system at the lower left (red arrow), for which the loss rates of the two resonators can then be written as:

\begin{equation}
    \label{equ:6-7}
    \gamma_1 = \frac{\mu_b^2}{2} + \frac{\mu_{\gamma 1}^2}{2} + \gamma_{int\_1}
\end{equation}
\begin{equation}
    \label{equ:6-8}
    \gamma_2 = \frac{\mu_c^2}{2} + \frac{\mu_{\gamma 2}^2}{2} + \gamma_{int\_2}
\end{equation}

Here, the intrinsic loss rates of the two resonators are represented by the terms, $\gamma_{int\_i}$ and the respective waveguide transmitted signals, at the three outputs can be written:
\begin{equation}
\left|t_{upper}\right|^2=\frac{\left|-j\mu_ca_2\right|^2}{\left|a_{in}\right|^2} \label{equ:6-9}
\end{equation}
\begin{equation}
\left|t_{middle}\right|^2=\frac{\left|-j\mu_{\gamma1}a_1-j\mu_{\gamma2}a_2\right|^2}{\left|a_{in}\right|^2} \label{equ:6-10}
\end{equation}
\begin{equation}
\left|t_{lower}\right|^2=\frac{\left|a_{in}-j\mu_ba_1\right|^2}{\left|a_{in}\right|^2} \label{equ:6-11}
\end{equation}

When $\gamma_{int\_i}=0$, the system is lossless and the conservation of energy then implies the total input power equals the total output power, yielding:
\begin{equation}
|t_{\text{upper}}|^2 + |t_{\text{middle}}|^2 + |t_{\text{lower}}|^2 = 1 \label{equ:6-12}
\end{equation}

Combining Eqs. \ref{equ:6-6}-\ref{equ:6-12}, the following relation is obtained.

\begin{equation}
\gamma_a = \frac{\mu_{\gamma 1} \, \mu_{\gamma 2}}{2} \label{equ:6-13}
\end{equation}

And substituting Eq. \ref{equ:6-13} into Eq. \ref{equ:6-6}, the TCMT equation for the tri-waveguide lossy-coupled MRR system can be written as:

\begin{equation}
j \frac{d}{dt}
\begin{bmatrix}
a_1 \\
a_2
\end{bmatrix}
=
\begin{bmatrix}
\omega_1 - j \gamma_1 & \mu_a - j \frac{\mu_{\gamma 1} \mu_{\gamma 2}}{2} \\
\mu_a - j \frac{\mu_{\gamma 1} \mu_{\gamma 2}}{2} & \omega_2 - j \gamma_2
\end{bmatrix}
\begin{bmatrix}
a_1 \\
a_2
\end{bmatrix}
+
\begin{bmatrix}
\mu_b \\
0
\end{bmatrix}
a_{in} \label{equ:6-14}
\end{equation}

From the perspective of classical electrodynamics, in this case, the real part of the coupling coefficient, $\mu_a$ represents direct photon exchange between the two resonators. By contrast, the imaginary part arises from indirect coupling mediated by the common central waveguide.

In this indirect process, light first couples from the lower resonator into the central waveguide, with a fraction of this coupling into the upper resonator and the rest coupling back into the lower resonator. This repeated exchange introduces an additional self-coupling contribution, causing the effective resonance frequency of each MRR to depend not only on its intrinsic geometry and material properties, but also on the coupling process itself.

It is important to emphasize that this theoretical description applies generally to inter-resonator coupling processes involving loss, regardless of whether this occurs via an intermediate waveguide or by radiation into free space.

\subsection{Eigenvalue analysis}

According to Eq. \ref{equ:6-14}, the Hamiltonian of the system shown in Fig. \ref{fig:6-1} is given by:

\begin{equation}
H
=
\begin{bmatrix}
\omega_1 - j \gamma_1 & \mu_a - j \frac{\mu_{\gamma 1} \mu_{\gamma 2}}{2} \\
\mu_a - j \frac{\mu_{\gamma 1} \mu_{\gamma 2}}{2} & \omega_2 - j \gamma_2
\end{bmatrix}
\label{equ:6-15}
\end{equation}

For two identical and symmetric MRRs, i.e., $\omega_1=\omega_2=\omega_0$ and $\gamma_1=\gamma_2=\gamma_0$, the eigenvalues of the Hamiltonian can be written as:

\begin{equation}
    \lambda_{1,2} = \omega_0 \pm \mu_a - j\left(\gamma_0 \pm \frac{\mu_{\gamma 1}\mu_{\gamma 2}}{2}\right)
    \label{equ:6-16}
\end{equation}

And, as before, the real (imaginary) parts of these eigenvalues represent the resonance frequencies (linewidths). Eq. \ref{equ:6-16} reveals that coupling-induced loss only modifies the linewidths of the resonances, but not their frequencies. In addition, it predicts a broader linewidth for the higher-frequency resonance.

In the special, dissipative coupling case where the coupling coefficient is purely imaginary, i.e., $\mu_a=0$, the resonances of the two MRRs are degenerate in the real part. However, as we will show from both numerical (Finite Element Method, FEM) simulations and experimental data, there is clearly a lifting of this degeneracy, most notably as the coupling rate increases, revealing separate (split) resonances, consistent with the lossy-coupled TCMT model developed here.

\subsection{Spectral response analysis}
In the steady-state regime, the coupled system is governed by:

\begin{equation}
    \frac{d a_i}{d t} = -j \omega a_i
    \label{equ:6-17}
\end{equation}

\begin{figure}[h]
    \centering
    \includegraphics[width=0.9\linewidth]{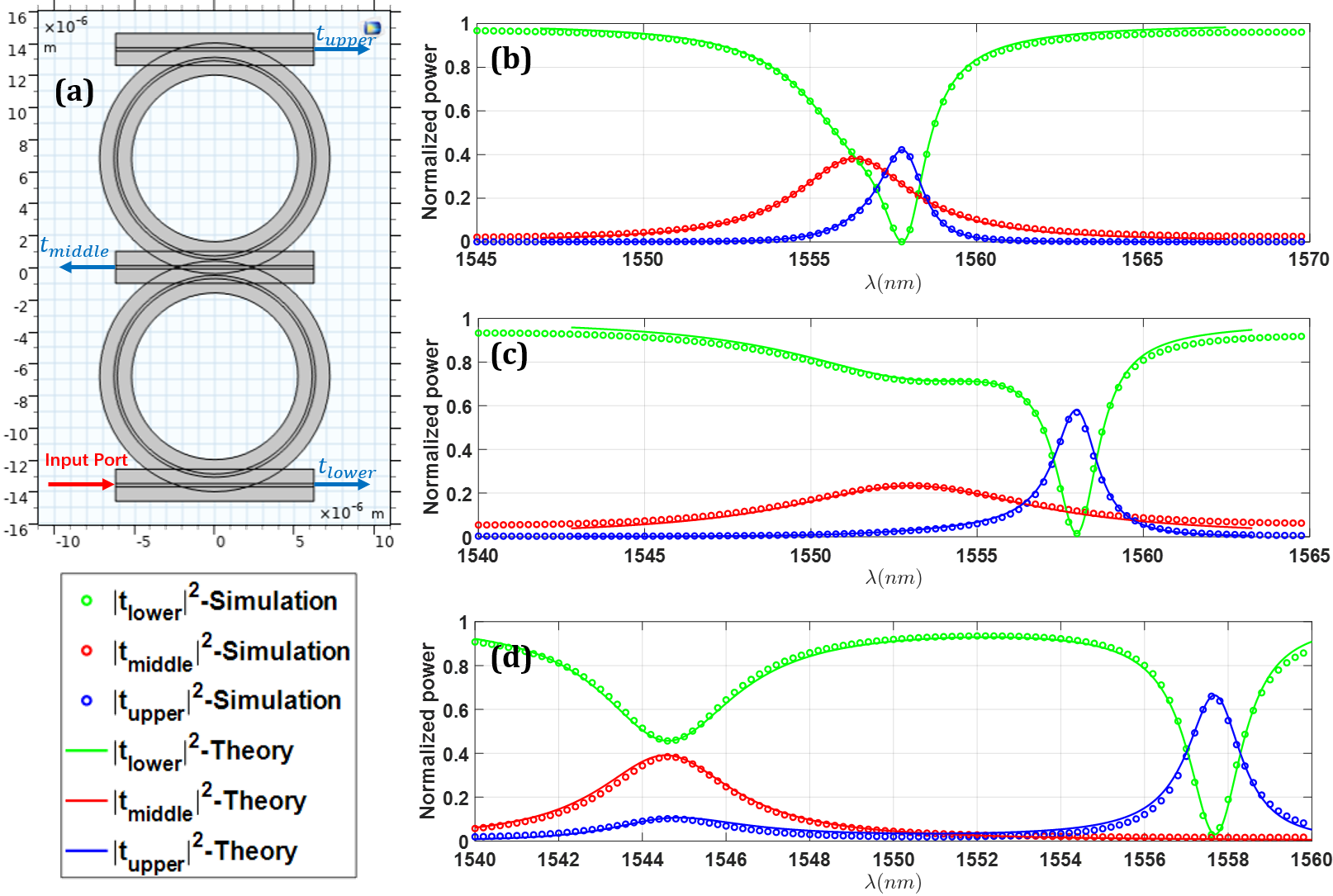}
    \caption{(a) The tri-waveguide `lossy' coupled MRR system built in COMSOL and comparison of the spectral response derived by FEM simulation (lines) with the theoretical model developed here (points) for each output of the system under the conditions of (b) low, (c) medium, and (d) high coupling rates, as per Table \ref{tab:6-1}.}
    \label{fig:6-2}
\end{figure}

By substituting Eq. \ref{equ:6-16} into Eq. \ref{equ:6-14}, the spectral response, Fig. \ref{fig:6-2} can be calculated for two identical MRRs, using the parameters listed in Table \ref{tab:6-1}, for three different coupling rates (low, medium and high).

\begin{table}[htbp] 
\centering
    \renewcommand{\arraystretch}{1}
\caption{Parameters used to derive theoretical responses in Figure \ref{fig:6-2} \label{tab:6-1}}
\begin{tabular}{cccc}
\hline
 &  Low coupling rate (THz)&  Medium coupling rate (THz)& High coupling rate (THz)\\
\hline
         $\nu_0=\omega_0/2\pi$&  192.53&  192.727& 193.276
\\
         $\mu_a$&  0.6083&  1.955 & 5.152 
\\
         ${\mu_{\gamma1}}^2$&  1.043 &  2.957 & 0.9452 
\\
         ${\mu_{\gamma2}}^2$&  1.043 &  2.957 & 0.9452
\\
         ${\mu_b}^2$&  1.005 &  1.005 & 1.005 
\\
         ${\mu_c}^2$&  1.005 &  1.005 & 1.005
\\
         $\gamma_1$&  1.134 &  2.092 & 1.086 
\\
         $\gamma_2$&  1.134 &  2.092 & 1.086 
\\
\hline
\end{tabular}
\end{table}

To validate our theoretical model, we simulated the transmission at the three output ports of the tri-waveguide coupled MRR system using the finite element method (FEM) tool, COMSOL Multiphysics. The COMSOL modelling utilises a 2D model; the model and domain size are shown in Fig. 2(a). The scattering boundary condition is applied, with the cladding width being 10 times that of the core width.

Fig. \ref{fig:6-2} reveals very good agreement between our model and the FEM simulation. Notably, as the coupling rate increases (coupling gap between MRRs and central waveguide decreases) (Fig. \ref{fig:6-2} (b) $\rightarrow$ (d)), the single resonance hybridizes into two distinct resonances that appear at the upper and lower outputs of the system. However, note that only a single resonance consistently appears at the common (middle) output, even for the highest coupling rate, Fig. \ref{fig:6-2} (d).

\begin{figure}[h]
    \centering
    \includegraphics[width=0.75\linewidth]{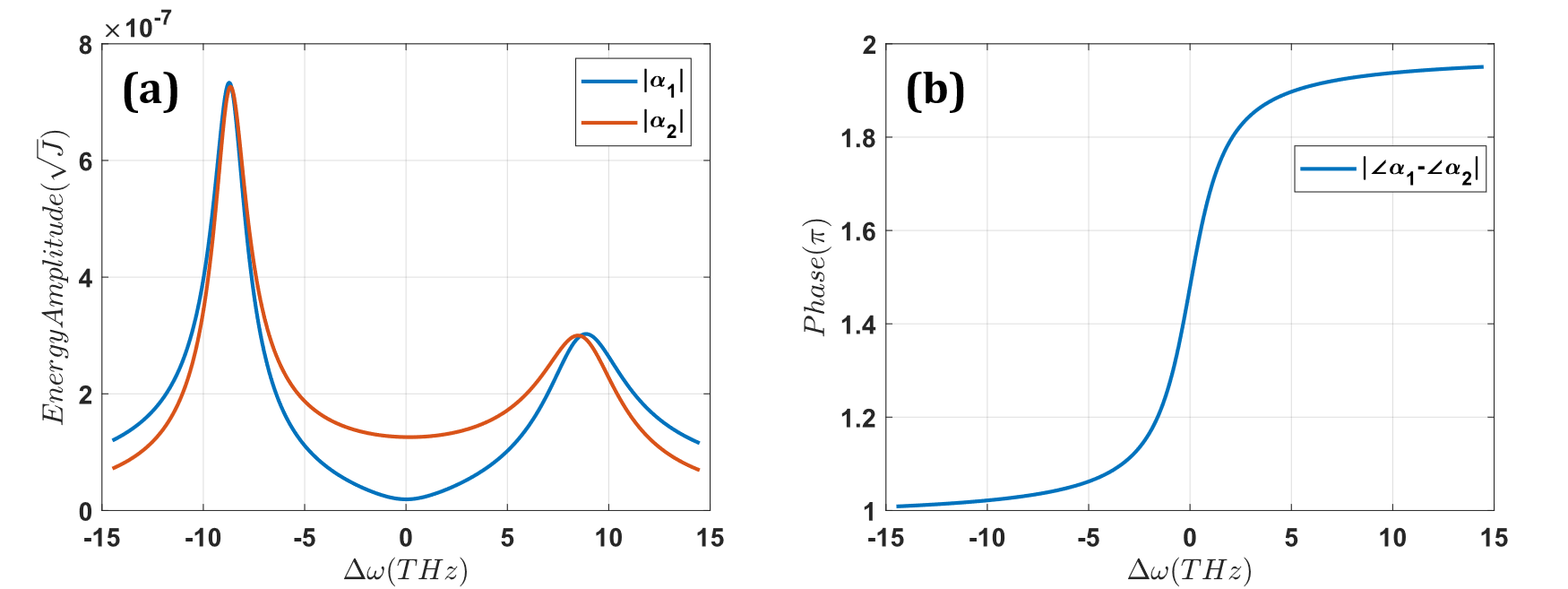}
    \caption{(a) Energy amplitude and (b) phase difference as a function of $\Delta \omega (=\omega-\omega_0)$ in the two MRRs at high coupling rate}
    \label{fig:6-3}
\end{figure}

This is the result of interference of the optical fields of the MRRs emerging at the middle output. Although the intracavity amplitudes of both resonators exhibit two resonances, these are almost completely out of phase, as confirmed in Fig. \ref{fig:6-3}.

\begin{figure}[h]
    \centering
    \includegraphics[width=0.75\linewidth]{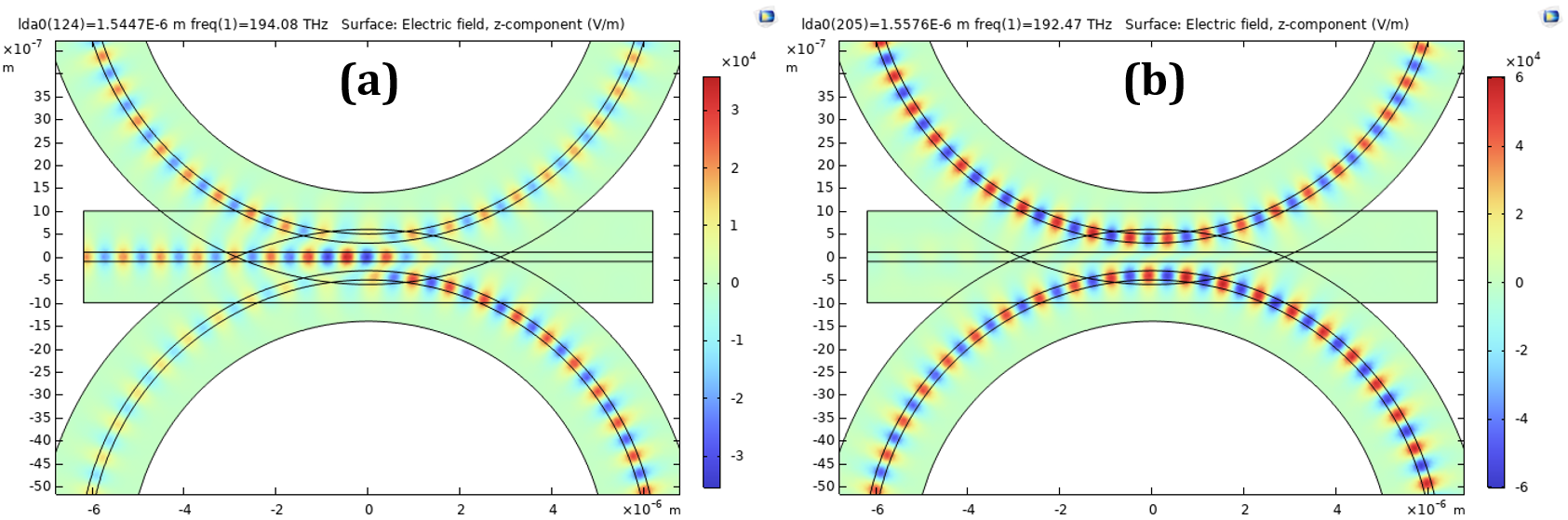}
    \caption{COMSOL simulated E-field distribution around the coupling region of the tri-waveguide coupled MRR system under the high coupling rate condition for the two specific resonances, (a) $\lambda=1544.7nm$ and (b) $λ=1557.6nm$ shown in Fig. \ref{fig:6-2} (c).}
    \label{fig:6-4}
\end{figure}

The COMSOL simulated E-field distribution around the coupling region, Fig. \ref{fig:6-4} provides further physical insight.

Fig. \ref{fig:6-4} reveals the longer-wavelength (1557.6 nm) resonance of the coupled MRR system to be an anti-symmetric (AS) mode for which the optical fields completely cancel at the common (middle) waveguide. On the other hand, the mode associated with the shorter-wavelength (1544.7 nm) resonance is symmetric (S), resulting in constructive interference and an observable output at the middle waveguide. Since this S-mode couples more strongly to the middle waveguide, it experiences a larger effective loss rate, which explains the relatively large linewidth of this resonance, consistent with that reported for a similar system in \cite{zhou_wavelength-selective_2015}.

\section{Design and fabrication of a novel, coherent `lossy coupled' MRR system}

\begin{figure}[h]
    \centering
    \includegraphics[width=0.9\linewidth]{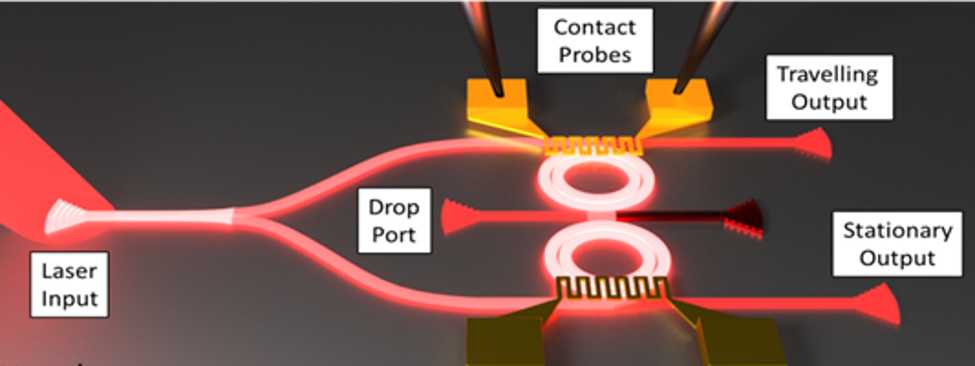}
    \caption{Schematic of the coherently driven tri-waveguide coupled MRR device. Light is coupled into the device from an optical fiber via a GC (far left), and divided by the y-splitter into the upper and lower branches and coupled to the MRRs or transmitted to the output GCs (far right). Lossy coupling of the resonant signals in the MRRs occurs at the common (middle) waveguide and hybridized S/AS modes collected at the drop port GC. Independent thermo-optic tuning of the relative phase and detuning is provided by the integrated TiN heaters via electrical contacts   }
    \label{fig:6-5}
\end{figure}

As we have shown that the output of the common (middle) inter-ring coupling waveguide serves as an effective `readout' of the S- and AS-super-modes of the coupled MRR system, we leveraged this principle by designing the coherently driven system shown schematically in Fig. \ref{fig:6-5}.

\begin{figure}[H]
    \centering
    \includegraphics[width=0.75\linewidth]{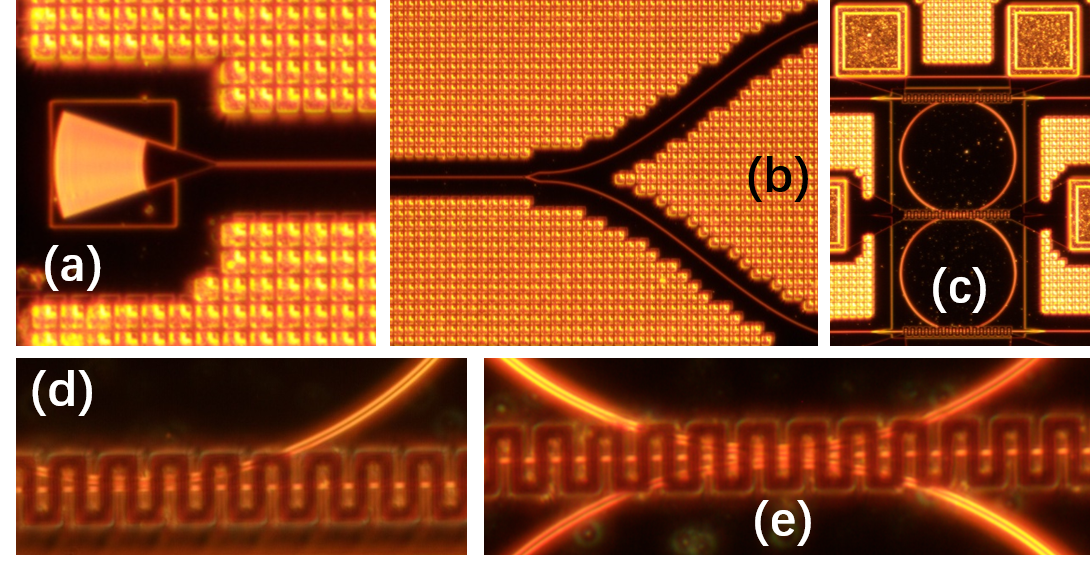}
    \caption{Optical micrographs of the different sections of our fabricated device showing (a) the input grating coupler (same design is used for output gratings), (b) the 3dB (y-) splitter, (c) the coupled MRR section with integrated TiN heaters and electrical contact pads, (d) close up image of the lower bus-to-ring coupling section and (e) ring-central bus-ring coupling section. Images are not to the same scale (for clarity).\label{fig:6-5+1}}
\end{figure}

The device in Fig. \ref{fig:6-5} was fabricated using the \textit{CMC Microsystems} silicon photonics foundry in Kingston, Ontario, Canada, using their Advanced Micro Foundry (AMF) general purpose fabrication process.  The device is patterned using a 193nm deep-UV lithography process from a 220 nm commercial silicon-on-insulator (SOI) wafer with a 3um buried oxide (BOX) to realise a 130 nm rib waveguide with a 90 nm silicon slab layer. Fig. \ref{fig:6-5+1} shows a series of optical micrographs of the key components of our device, including input grating coupler, 3dB (y-) splitter, lower bus-to-ring and ring-central bus-ring coupling sections (plus zoomed in images of both), integrated TiN heaters and electrical contact pads.

Light is fiber-coupled to the device using a grating coupler (GC), Fig. \ref{fig:6-5+1} (a), designed for a central operating wavelength of 1550 nm, with an etch (groove) depth of 70 nm, and 24 periods of 630 nm length. Single transverse electric (TE) mode propagation is achieved with a waveguide width, $w$ = 320 nm, as we confirmed using a 2D lumerical simulation (not shown). A 3 dB (Y-) splitter, Fig. \ref{fig:6-5+1} (b) divides the propagating mode into the upper and lower branches, where it is either resonantly coupled to the identical (by design) MRRs, across a 300 nm coupling gap, or transmitted directly to the `Stationary' or `Travelling' outputs, again via GCs of the same design as the input. The MRRs, Fig. \ref{fig:6-5+1} (c) are comprised of concentric (slot) waveguides, enabling lower loss (higher Q-factor) with an inner waveguide width, $w_{inner}$ = 250 nm, outer waveguide width, $w_{outer}$ = 290 nm and slot width, $w_{slot}$ = 200 nm. The MRR radius, measured to the centre of the slot, $r_{MRR}$ = 25 $\mu m$. The common (middle) coupling waveguide, width, $w_{middle}$ = 320 nm is also terminated with the same GC design, to enable collection of the combined `Drop' output signal(s). The device was capped with a 1.3 $\mu m$ SiO\textsubscript{2} layer before depositing TiN resistive heater elements, Fig. \ref{fig:6-5+1} (d, e) and associated electrical contact pads above the coupling section of each MRR, providing independent (relative) detuning of their cavity resonance condition, and a relative phase shift, $\varphi$, via the thermo-optic effect.

\begin{figure}[H]
    \centering
    \includegraphics[width=0.75\linewidth]{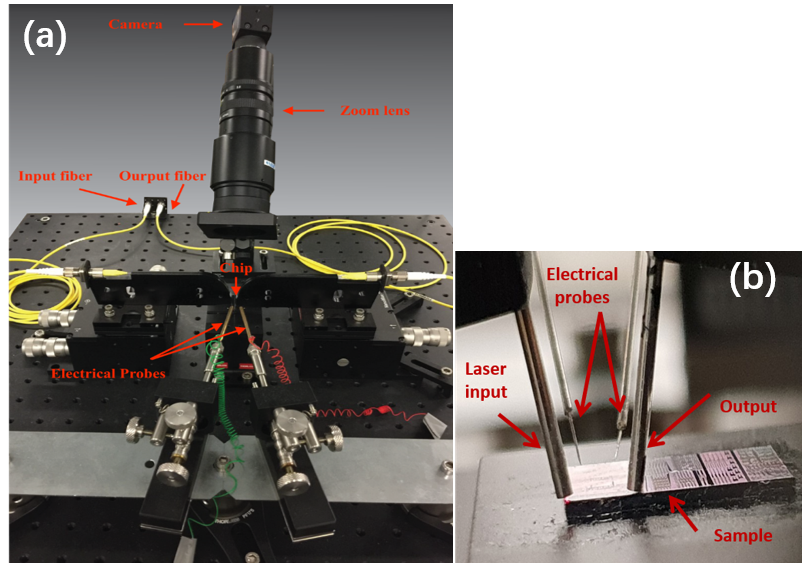}
    \caption{Fiber-to-chip measurement set-up including extended InGaAs (0.4-1.7um) camera and zoom lens (up to 24.5x magnification), input and output piezo-controlled fiber positioning stages and electrical contact probes for thermo-optic tuning. (b) Close-up of the delivery and collection of light to/from the silicon photonics chip using optical fibers and electrical contact probes for thermo-optic tuning.}
    \label{fig:6-5+2}
\end{figure}

Fig. \ref{fig:6-5+2} shows images of our fiber-to-chip experimental measurement set-up, including a close-up of the silicon photonics chip in-situ, with input and output optical fibers and electrical probes.

Keeping with the same nomenclature as in Section 3, the temporal coupled-mode equation describing this system is:

\begin{equation}
j \frac{d}{dt}
\begin{bmatrix}
a_1 \\
a_2
\end{bmatrix}
=
\begin{bmatrix}
\omega_1 - j\gamma_1 & \mu_a - j \frac{\mu_{\gamma 1}\mu_{\gamma 2}}{2} \\
\mu_a - j \frac{\mu_{\gamma 1}\mu_{\gamma 2}}{2} & \omega_2 - j\gamma_2
\end{bmatrix}
\begin{bmatrix}
a_1 \\
a_2
\end{bmatrix}
+
\begin{bmatrix}
\mu_b \\
\mu_c e^{i\phi}
\end{bmatrix}
\frac{a_{\text{in}}}{\sqrt{2}}
\label{equ:6-18}
\end{equation}

And the transmitted intensity at the three output ports is:

\begin{equation}
\left| t_{\text{stationary}} \right|^2 
= \frac{\left| \frac{a_{\text{in}}}{\sqrt{2}} - j \mu_b a_1 \right|^2}{\left| a_{\text{in}} \right|^2} \label{equ:6-19}
\end{equation}
\begin{equation}
\left| t_{\text{drop}} \right|^2 
= \frac{\left| -j \mu_{\gamma 1} a_1 - j \mu_{\gamma 2} a_2 \right|^2}{\left| a_{\text{in}} \right|^2}\label{equ:6-20}
\end{equation}
\begin{equation}
\left| t_{\text{travelling}} \right|^2 
= \frac{\left| \frac{a_{\text{in}}}{\sqrt{2}} e^{i\phi} - j \mu_c a_2 \right|^2}{\left| a_{\text{in}} \right|^2}
\label{equ:6-21}
\end{equation}

\subsection{Zero detuning}
We first consider the case in which the two MRRs have identical resonance frequencies, $\omega_1=\omega_2$ and we examine the effect of any phase shift, $\varphi$, on the spectral response of the system.

\begin{figure}[H]
    \centering
    \makebox[\textwidth][c]{\includegraphics[width=1.4\linewidth]{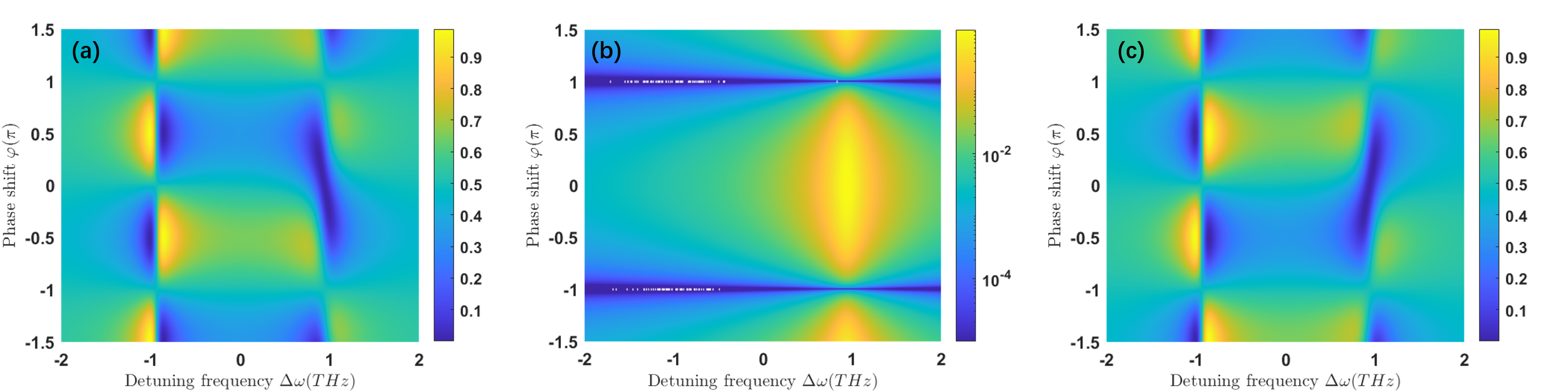}}
    \caption{Spectral intensity maps showing the split resonances for the (a) Stationary, (b) Drop and (c) Travelling ports, as a function of the phase shift $\varphi$ with zero detuning, i.e., for $\omega_1=\omega_2$.}
    \label{fig:6-6}
\end{figure}

Using the parameters listed in Table \ref{tab:6-2}, the spectral responses shown in Fig. \ref{fig:6-6} are obtained.

\begin{table}[H] 
\centering
\caption{Parameters used to derive theoretical responses in Fig. \ref{fig:6-6} \label{tab:6-2}}
\renewcommand{\arraystretch}{1}
\begin{tabular}{cc}
\hline
Parameter &  Coupling rate  (GHz)\\
\hline
         $\mu_a$&  940.4 
\\
         ${\mu_{\gamma1}}^2$&  58.78 
\\
         ${\mu_{\gamma2}}^2$&  58.78 
\\
         ${\mu_b}^2$&  150.5 
\\
         ${\mu_c}^2$&  150.5 
\\
         $\gamma_1$&  105.4
\\
         $\gamma_2$&  105.4  
\\
\hline
\end{tabular}
\end{table}

Here we use a relatively large inter-ring coupling rate, $\mu_a$ to ensure S and AS mode far enough away to observe the complete waveform corresponding to each mode and examine the resonance splitting only as a function of the relative phase, $\varphi$.

The drop-port spectrum resembles that of the lossy-coupled resonator system discussed earlier, whereas the stationary and travelling ports exhibit Fano-like spectral line shapes arising from coherent interference.

\begin{figure}[H]
    \centering
    \includegraphics[width=0.8\linewidth]{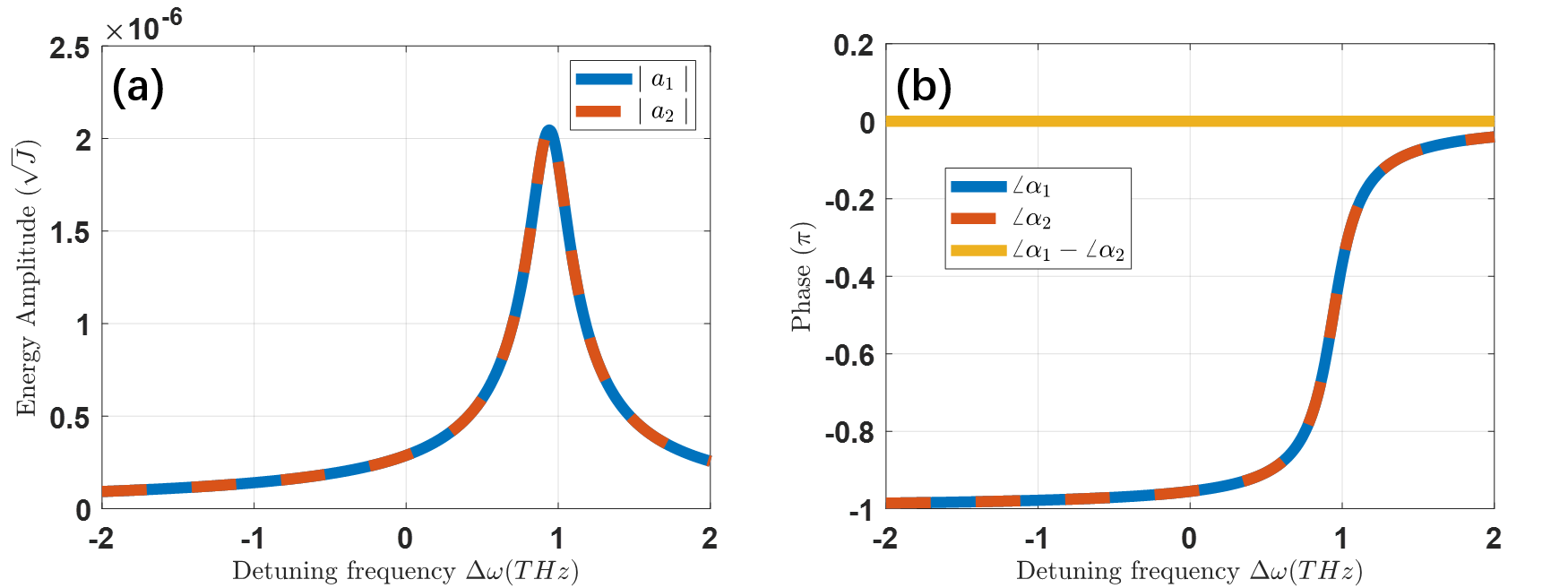}
    \caption{(a) Energy amplitude and (b) phase in the two MRRs when the relative phase shift, $\varphi=0$}
    \label{fig:6-7}
\end{figure}

When $\varphi = 0$, the optical mode in the upper and lower waveguides are identical. Under this condition, the AS mode is suppressed, and only the S mode appears, shifted from the MRR resonant frequency by an amount, $\Delta \omega \approx 1 THz$, Fig. \ref{fig:6-7} (a).

The fact that the optical fields in the two MRRs are in phase, Fig. \ref{fig:6-7} (b) leads to the appearance of the S mode at the output of the common (middle) waveguide.

\begin{figure}[H]
    \centering
    \includegraphics[width=0.8\linewidth]{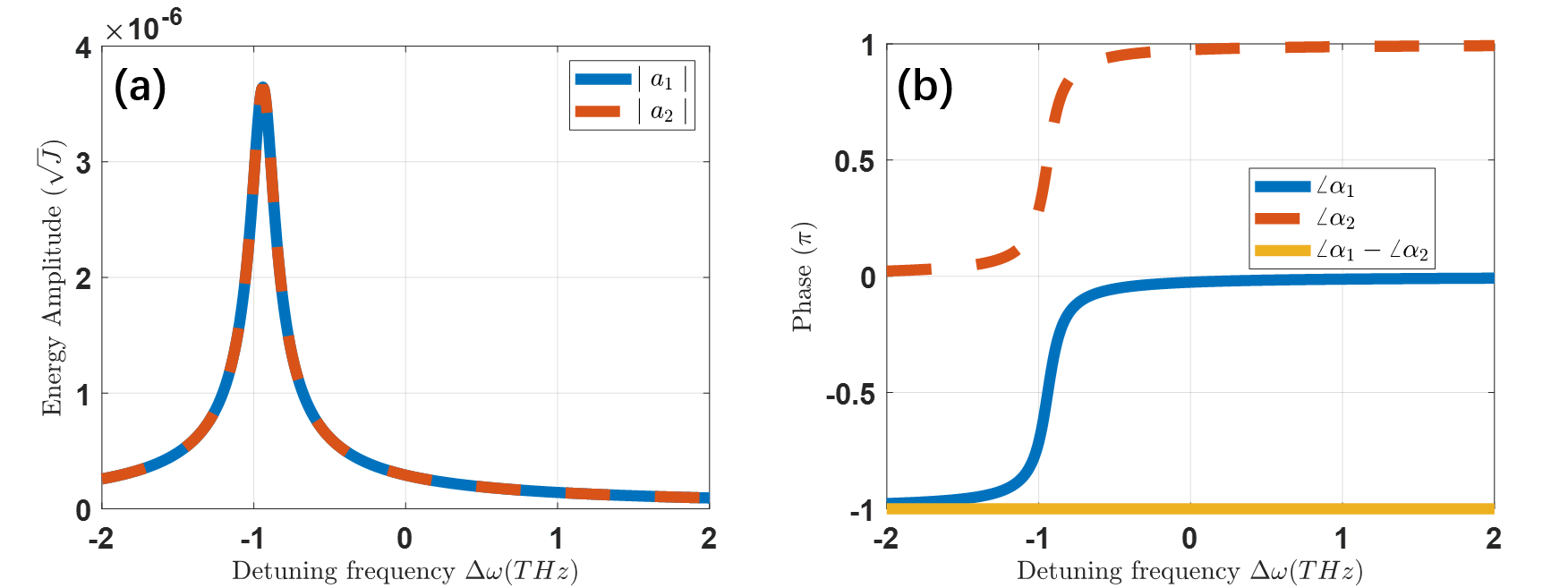}
    \caption{(a) Energy amplitude and (b) phase in the two MRRs when the relative phase shift, $\varphi = \pi$}
    \label{fig:6-8}
\end{figure}

However, for a relative phase, $\varphi = \pi$, the S mode is suppressed, and only the AS mode appears, in this case shifted from the MRR resonant frequency by an amount, $\Delta \omega \approx -1 THz$, Fig. \ref{fig:6-8} (a).

The relative phase, $\varphi = \pi$ between the optical fields in the two MRRs, Fig. \ref{fig:6-8} (b) explains the destructive interference (and zero output) at the common drop port.

\subsection{Small relative detuning}

\begin{figure}[H]
    \centering
    \makebox[\textwidth][c]{\includegraphics[width=1.4\linewidth]{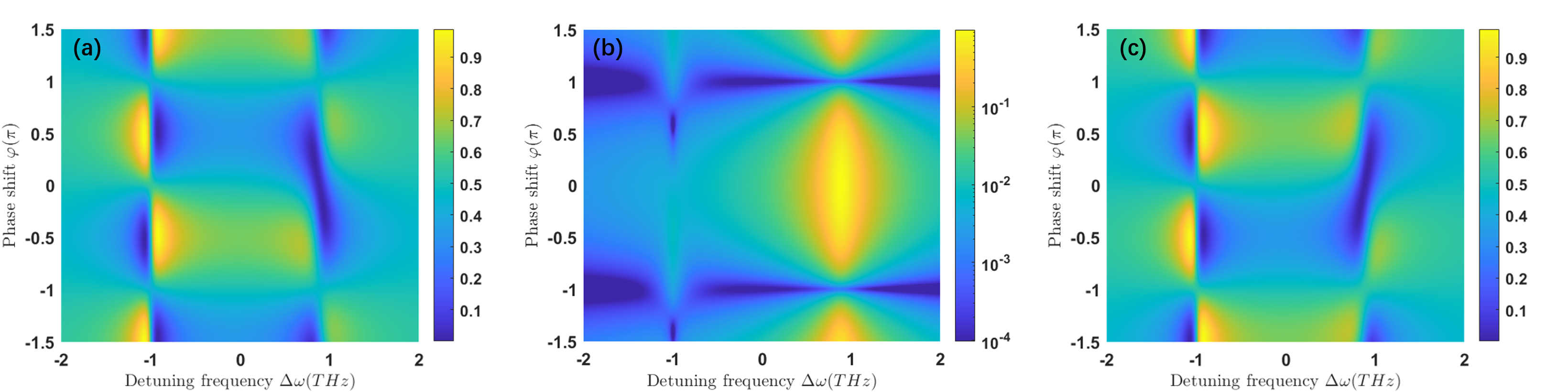}}
    \caption{Spectral intensity maps showing the split (hybridized) resonances for the (a) Stationary, (b) Drop and (c) Travelling ports, as a function of the phase shift $\varphi$ with a small (fixed) relative detuning, i.e., for $\Delta \omega=\omega_1-\omega_2= 117.5 GHz$}
    \label{fig:6-9}
\end{figure}

In practice, fabrication imperfections make it difficult to ensure perfectly identical MRRs, and thus resonance frequencies. We therefore consider the effects of a small, fixed relative detuning while keeping all other parameters unchanged. Arbitrarily setting this to $\Delta \omega=\omega_1-\omega_2= 117.5 GHz$, we leave all other parameters as in Table \ref{tab:6-2} and show in Fig. \ref{fig:6-9} the effect this has on the spectral response of the system, again as a function of the phase shift, $\varphi$.

Compared with the zero-detuning case, the spectral responses at the `Stationary' and `Travelling' ports remain qualitatively similar. However, for the drop port, in addition to the S-mode related resonance (near +1 THz), a second, much narrower, resonance, associated with the AS mode, now appears (near -1 THz). Depending on the precise phase shift, this AS-mode related resonance can exhibit either an absorption, e.g., 
at $\varphi = \pi/2$ or a transparency, e.g., at $\varphi = -\pi/2$. Notably, the precise phase shift, $\varphi$, corresponding to the absorption peak gradually approaches zero as detuning increases. 

Although the Hamiltonian of Eq. \ref{equ:6-18} does not explicitly feature the phase, $\varphi$, and thus the eigenvalues should be independent of it, the coherent interference effects modify the observable spectral linewidths such that they may no longer be described solely by the imaginary parts of the Hamiltonian eigenvalues.

This novel device may therefore be used to realize absorption resonance, on demand, when there exists a small (fixed) relative detuning between the two MRRs (as there almost always is in such fabricated PIC architectures), simply by controlling the relative phase, $\varphi$. In the following section, we reveal this control, via thermo-optic tuning, on the measured spectral characteristics at the drop port of our fabricated device. It should be noted, however, that the relatively high inter-ring coupling rate assumed in the previous (model) analysis is rather difficult to achieve in practice, meaning the observed eigen-peaks are experimentally much closer together in our prototype device.

\section{Experimental results and discussion}

\begin{figure}[H]
    \centering
    \makebox[\textwidth][c]{\includegraphics[width=1.4\linewidth,
    trim=0cm 1.1cm 0cm 1.3cm,
    clip]{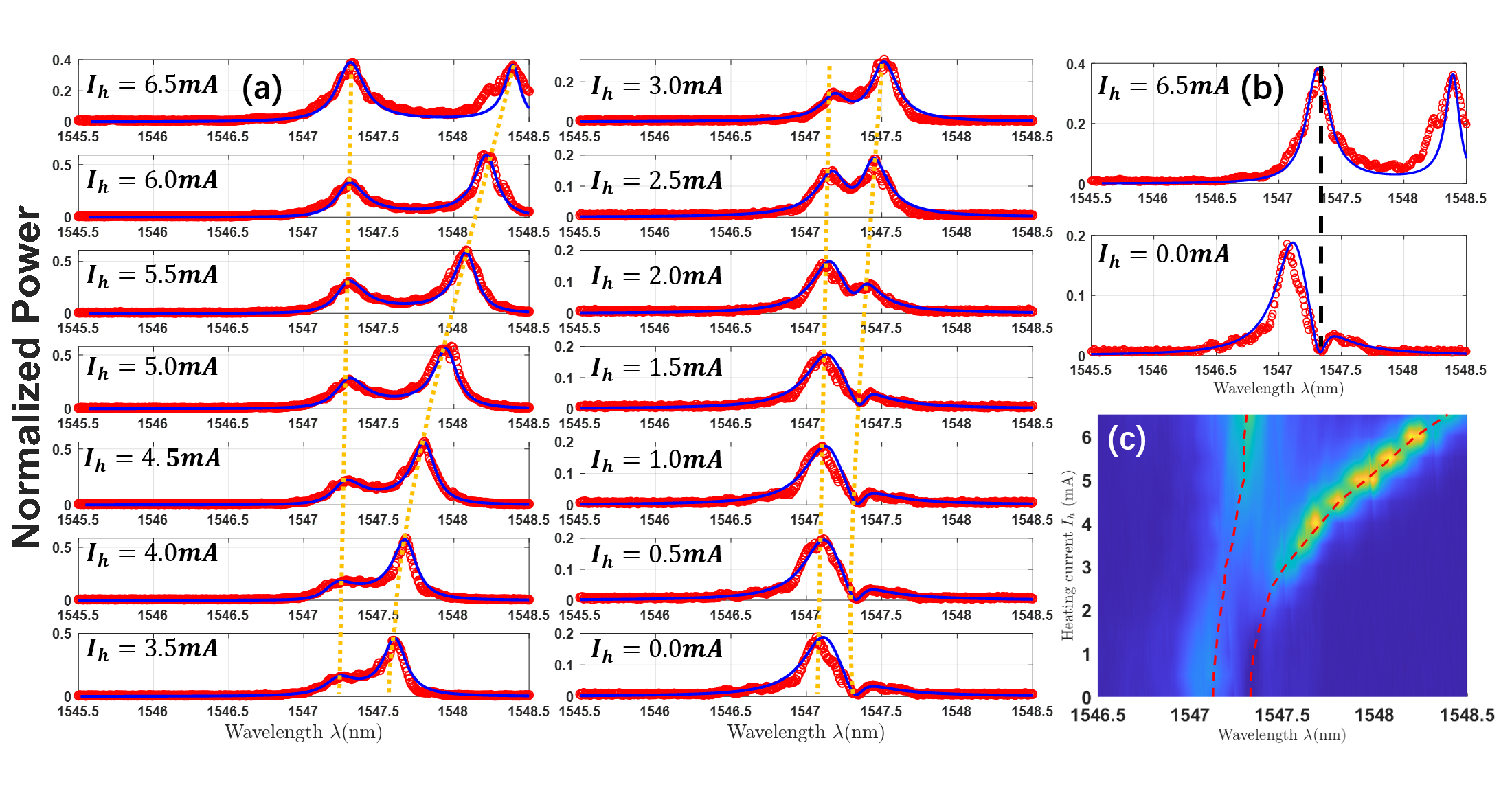}}
    \caption{(a) Measured (red points) and TCMT model fit (blue lines), using Eq. (18) from the drop port as a function of heater current in the range $0 mA \geq I_{heater} \geq 6.5 mA$. The yellow dashed lines show the approximate resonance positions as a guide to the eye (b) Comparison of the drop port spectra at $I_{heater} = 0 mA$ (near zero detuning) and $6.5 mA$ (maximum detuning) reveals the stationary resonance flipping between absorption resonance peak and transparency (c) Heat map of the measured spectral intensity and model eigenvalues (red dashed lines) describing the resonance positions, as a function of $I_{heater}$ revealing an `avoided crossing-like' behaviour that gives rise to the tunable absorption resonance.}
    \label{fig:6-10}
\end{figure}

Figure \ref{fig:6-10} shows the measured drop-port spectra from our fabricated device, along with the fitted curves, using Eq. \ref{equ:6-18}, for a heater current range, $0 mA \geq I_{heater} \geq 6.5 mA$. 

We note that, owing to the positioning of the integrated TiN heater(s) above the bus-to-ring coupling gap in our device, both the resonant frequency and coupling rate will be affected by the heating. As such, we have taken this into account in our model, by adjusting ${\omega }_{1}$ and ${\mu }_{b}$ when fitting our model to our measured data. Fig. \ref{fig:6-10} reveals that, as the heating current increases, the longer-wavelength resonance evolves continuously from an absorption `notch' to a transparency `peak'. The measured spectra show good qualitative agreement with the TCMT model across the full range of heater currents we have applied. Critically, they reveal that controllable  phase-controlled super-mode switching of absorption can be achieved. While the spectral linewidths can no longer be directly described by the imaginary parts of the Hamiltonian eigenvalues, due to the coherent interference, the red dashed lines in Fig. \ref{fig:6-10} (c) clearly show that the resonance peak positions are faithfully reproduced by the real parts of the Hamiltonian eigenvalues, from Eq. \ref{equ:6-18}. The avoidance of resonance crossover also demonstrates that the real part of the coupling rate for this system is non-zero.

\begin{figure}[h]
    \centering
    \includegraphics[width=0.8\linewidth]{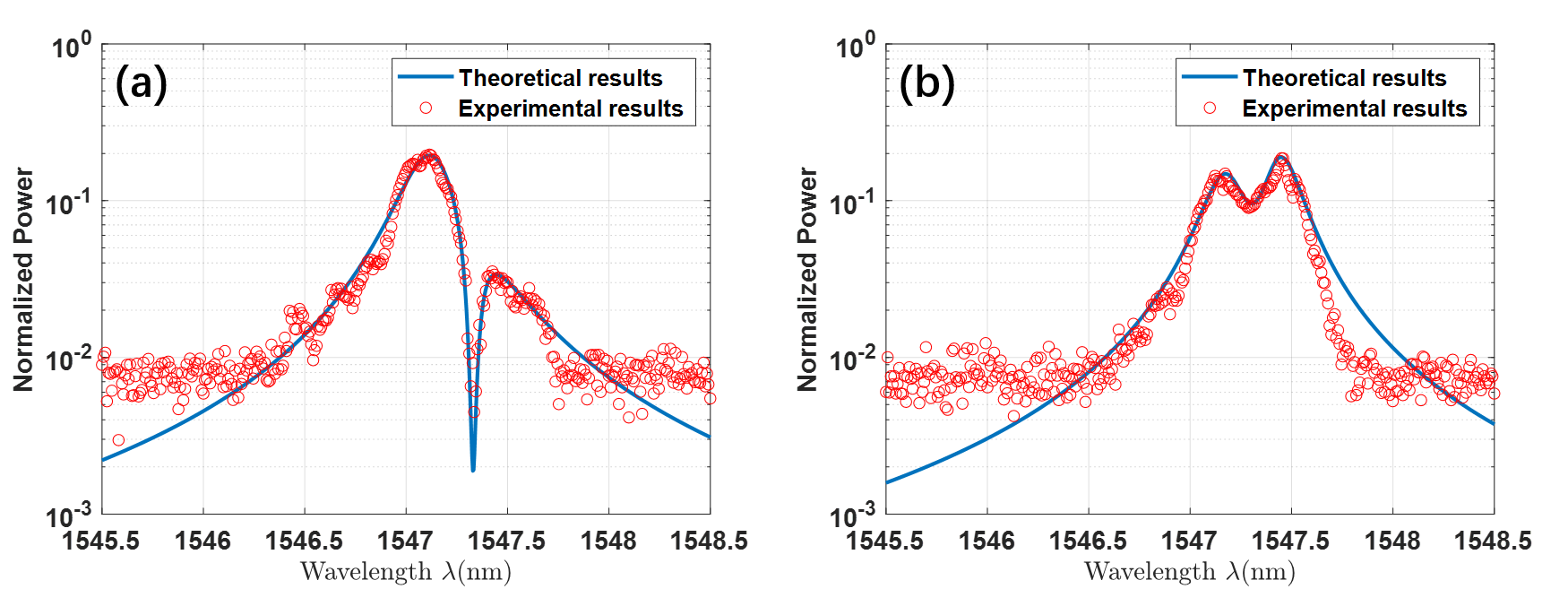}
    \caption{Measured (red points) and fitted (blue lines) using Eq. \ref{equ:6-18} Drop port spectra for (a) $I_{heater}=0.5mA$ and (b) $I_{heater}=2.5mA$. Departure of the model from the measured data away from the resonance(s) is due to the instrument (optical spectrum analyser) noise floor.}
    \label{fig:6-11}
\end{figure}

Finally, in Fig. \ref{fig:6-11} we show two representative drop port spectra, for $I_{heater}=0.5mA$ and $I_{heater}=2.5mA$, that highlight the fact that thermal tuning in this device simultaneously alters the spectral position and linewidth, i.e., by changing both the relative phase and detuning. We attribute this to the choice of positioning of the integrated heaters in our device, directly above the coupling region from the outer waveguide to the MRRs. 

The spectral resonance peaks and linewidths in Fig. \ref{fig:6-11} are well described by our TCMT model using Eq. \ref{equ:6-18}. This gives confidence in our mathematical description of the device, and should allow for optimization of such phase-controlled super-mode driven absorption sources for a broad range of integrated photonics-based applications.

\section{Conclusion}
In this work, we introduced the concept of `lossy coupling' in a tri-waveguide coupled MRR system and derived a TCMT framework incorporating this coupling-induced loss. We showed that `lossy coupling' in such structures, via a common central waveguide allows for selective readout of the S/AS hybrid resonance modes. Building on this principle, we proposed and fabricated a coherently driven lossy-coupled MRR device in the silicon photonics platform to explore the effects of phase and relative detuning on the common drop port transmission spectra, using integrated heaters. Our TCMT model of this device predicts phase-controlled S and AS super-modes, and thus tunable, or `on-demand' absorption resonances. Experimental data confirms these predictions and demonstrates that the proposed platform can provide a practical route for tunable photonic functionality, with potential applications in optical switching, sensing, and integrated coherent control.

\begin{backmatter}

\bmsection{Acknowledgment}
One of us (T.Li) wishes to thank the University of Manchester for the award of a doctoral scholarship. This work was supported by the Henry Royce Institute for Advanced Materials, funded through EPSRC grants EP/R00661X/1, EP/S019367/1, EP/P025021/1, and EP/P025498/1, and the EPSRC Centre for Doctoral Training (CDT) in Compound Semiconductor Manufacturing, under grant code EP/Y035801/1.

\bmsection{Disclosures}
The authors declare no conflicts of interest.

\end{backmatter}



\bibliography{references}






\end{document}